\documentclass[a4paper,12pt,prb]{revtex4}
\usepackage{amsmath,graphicx}
\newcommand{\Sp}{\operatorname{Sp}}
\begin{document}
\title{
KINETICS OF 2D ELECTRONS IN THE MAGNETIC FIELD IN THE PRESENCE OF
MICROWAVE RADIATION. RESPONSE OF A NON-EQUILIBRIUM SYSTEM TO A
WEAK MEASUREMENT FIELD}
\author{I. I. Lyapilin}
\email{Lyapilin@imp.uran.ru}
\author{A. E. Patrakov}
\affiliation{Institute of Metal Physics, UD of RAS, Yekaterinburg,
Russia}

\begin{abstract}
Kinetics of spatially uniform distribution of 2D electrons in crossed
electric and magnetic fields in the presence of microwave radiation has been
studied. In the present model the contribution from the microwave radiation
and the effects of Landau quantization are considered exactly, while
scattering is treated perturbatively. Here Landau---Floquet states interact
with the impurity potential that causes transitions between them. The linear
response of a non-equilibrium system to a weak measurement field has been
considered for the case when the non-equilibrium state of the carriers can
be described by the average values of the total energy and the number of
particles.
\end{abstract}

\maketitle
\section{Introduction}
In two-dimensional electron systems (2DES) with high
($\sim 10^7\ \mathrm{cm^2/Vs}$) mobility, the magnetoresistance exhibits
strong oscillations in pre-Shubnikov interval of magnetic fields under
microwave irradiation \cite{Zudov03,Mani02}. The parameter that governs
those oscillations is the ratio of the radiation frequency $\omega$ to the
cyclotron frequency $\omega_c$. The maxima of the magnetoresistance are
observed at $\omega/\omega_c = j - 1/4$, where $j$ takes integer values. The
resistance of the sample is minimal at $\omega/\omega_c = j + 1/4$, and with
sufficiently intensive radiation zero-resistance states are observed.

Let's mention the important features of the above-mentioned experiments. The
effect is observed under the conditions:
$\hbar/\tau\ll T \simeq \hbar\omega_c \leq \hbar\omega \ll\zeta$, from which
one can conclude that its nature is quasiclassical. Here $\tau$ is the
momentum relaxation time, $\omega \geq \omega_c$ are the radiation frequency
and the cyclotron frequency, respectively, $\zeta$ is the Fermi energy, $T$ is
the temperature expressed in units of energy.

Currently, there exist a sufficiently large number of theoretical papers
(\cite{Durst03,Shi03,Mikhailov03,Dmitriev0310,Dmitriev0304,Volkov03}, see
also \cite{Vavilov04} and references therein)
in which various aspects of the observed phenomena are considered. It is
necessary to note that the possibility of the existence of
negative-resistance states was first predicted in \cite{Ryzhii68}. Some of
the models are based upon the influence of the microwave radiation upon the
processes of scattering an electron along the weak dc electric field or in the
opposite direction. There are also alternative explanations that relate the
observed phenomena with non-trivial dependence of the non-equilibrium
distribution function upon the energy, that acquires oscillating character
under the effect of the microwave radiation. The schemes described above
presume that only bulk effects are responsible for the observed phenomena.
Other models are possible. E.g., the scenario proposed in \cite{Mikhailov03}
says that two mechanisms are actually important. One of them leads to a
resonance at the magnetoplasmon frequency and is of the bulk origin. The
second one is related to the development of drift plasma instability at the
edge of the sample. Both the proposed models and the predictions that follow
from them still require additional experimental validation.

In this paper, in order to explain the observed conductivity oscillations,
we present a model that incorporates contributions from
Landau quantization and microwave irradiation (in long-wavelength limit)
exactly, without the use of perturbation theory. Electron scattering upon
impurities is considered perturbatively. With respect to Landau---Floquet
states, impurities act as a coherent oscillating field that causes
transitions which are essential for reproducing the oscillating
magnetoresistance behavior. This problem is a classical variant of the
theory of non-equilibrium system response to a weak measuring field.
Indeed, under microwave irradiation, in the system under consideration a
non-equilibrium state is formed. The task is to find the response of such
possibly strongly non-equilibrium system to a weak measuring field.

It is worth mentioning that the theory of linear reaction upon external
mechanical perturbations is currently well-established.
Within this approach, the kinetic coefficients are expressed in terms of the
equilibrium correlation functions.
However, the situation changes radically if
it is required to find the response of a system that is already out of the
equilibrium, to an additional measurement field.
Problems of this kind are usually solved with the help of the method of
kinetic equations. We shall apply the theory of linear response of
a non-equilibrium system to a weak measurement field for analysis of the
transport phenomena for 2D charge carriers under microwave irradiation.

Our results depict strong oscillations of $\rho_{x x}$ with negative
resistance states. It has been found that that the microwave-induced
correction to $\rho_{x x}$ vanishes when
$\epsilon = \omega/\omega_c = j$, $j$ is an integer.
Oscillations have minima at $\epsilon = j + \delta$ and maxima at
$\epsilon = j - \delta$, where $\delta \approx 1/5$. The calculations have
been carried out for various polarizations of microwave electric field.
The role of the mobility of the carriers for the magnetoresistance
oscillations has been studied. It has been shown that magnetoresistance
oscillations should still be observable in the lower-mobility samples if one
increases the microwave radiation frequency.
A role of  multiphoton process in this phenomenon can also be relatively
easily studied using the described approach.

The article is organized in the following way. In section \ref{cantrans},
the canonical transformation of the Hamiltonian is considered. This allows
us to avoid the direct analysis of the system's reaction upon the ac
electric field of the microwave radiation that is not necessarily a weak
perturbation. Section \ref{linresp} contains a short introduction into the
theory of the linear response of a non-equilibrium system. This theory is
applied to the canonically-transformed system. In the same section, the main
results for the magnetoconductivity tensor are expressed. Finally, the
results of the numerical analysis are presented in section \ref{numanal},
followed by the conclusion.

\section{Canonical transformation}\label{cantrans}

Let's consider a system of 2D charge carriers and scatterers in a conducting
crystal under effect of dc magnetic $(0,0,B)$, dc electric $\vec{E} = (E_x,
0, 0)$ and ac electric $\vec{E}(t) = (E_x(t),E_y(t),0)$ fields. The
Hamiltonian of such system is can be written in the following form:

\begin{equation}\label{g1}
    H(t) = H_e + H_v + H_{ev} + H_{ef}(t)+ H^0_{ef},
\end{equation}
where
\begin{equation}\label{g2}
H_e = \sum_i\frac{P_i^2}{2m}=\sum_i\frac{(\vec{p}_i-
(e/c)\vec{A}_i)^2}{2m} = \sum_i\frac{1}{4m}(P^+_i P^-_i+P^-_i P^+_i)
\end{equation}
is the Hamiltonian of free charge carriers in the magnetic field,
$m$ is the effective mass, $e$ is the electron charge,
$P^\alpha_i$ is a component of kinetic momentum of electrons, and
$$
P^\pm = P_x \pm i P_y,\qquad [P^+, P^-]=2m\hbar\omega_c,\quad
[P^\pm, H_e]=\pm\hbar\omega_c P^\pm.$$
$\omega_c=|e| B / m c$ is the cyclotron frequency of electrons.
$$\vec{A}=-\frac{1}{2}[\vec{r}\times\vec{B}]$$
is the vector potential. $H_v$ is the Hamiltonian of the lattice, $H_{ev}$
is the Hamiltonian of the interaction of electrons with scatterers (in this
case, impurities are considered).  $H_{ef}(t)$ is the Hamiltonian of
interaction of conducting electrons with the ac electric field,
and $H^0_{ef}$ describes the interaction
with the external dc electric field $\vec{E}$ response to
which is what we are interested in.
\begin{eqnarray}\label{g23}
H^0_{ef} = -e E^\alpha \sum_i x^\alpha_i.
\end{eqnarray}

In the laboratory frame of reference, evolution of the state of a
free 2D electron is described by Schr\"odinger equation:
\begin{equation}\label{g3}
\{i\frac{\partial}{\partial t} - H_0(t)\} \psi(\vec{r},t) =
\{i\frac{\partial}{\partial
t} - H_0 + H_{ef}(t)\} \psi(\vec{r},t) = 0.
\end{equation}
Here $H_0(t)$ is the Hamiltonian of an electron without interaction with
scatterers, $\psi(\vec{r},t)$ is the wave function describing the electron
state. Let's perform a time-dependent canonical transformation $W_t$ that would
exclude the ac electric field from the Schr\"odinher equation, i. e.
\begin{equation}\label{g4}
W^\dagger_t\{i\frac{\partial}{\partial t} - H_0(t)\} W_t =
i\frac{\partial}{\partial t} - H_0.
\end{equation}

If such a transformation is done, then, obviously, any solution of Eq.
(\ref{g3}) can be represented as a result of canonical transformation of the
corresponding solution without the ac electric field, i. e.
\begin{equation}\label{g5}
\Psi(\vec{r},t) = W_t \psi(\vec{r},t).
\end{equation}
The unitary transformation of the state vector (\ref{g5}) is in fact a
change of the reference frame from the laboratory one to the one moving with
the time-dependent velocity
${\dot{\vec{\xi}}(t)}=\partial \vec{\xi}/\partial t$. Thus, it corresponds
to the shift of coordinates
$\vec{r}\rightarrow\vec{r}+\vec{\xi}(t)$ and momenta
$\vec{p}\rightarrow\vec{p}+m{\dot{\vec{\xi}}(t)}$.

Note that, in the presence of magnetic field, the translation in the
coordinate space should be considered as the operation of ``magnetic
translation'', defined as
\begin{equation}\label{g6}
T_{\vec{\xi}}\Psi(\vec{r})=\exp\{i\frac{e}{c}
\vec{A}(\vec{\xi})\cdot\vec{r}\}
\exp(-i\vec{\xi}\cdot\vec{p})\Psi(\vec{r}),
\end{equation}
where $\vec{A}(\vec{\xi})=(\vec{B}\times\vec{\xi})/2$ is the vector
potential of the external magnetic field in the symmetric gauge:
$\vec{A} = (-\frac{1}{2} y B, \frac{1}{2} x B, 0)$.
The operator $W_t$ that defines the unitary (canonical) transformation can
be represented in the following form:
\begin{equation}\label{g7}
W_t = \exp\{i\theta(t)\}\exp\{-i\vec{\xi}\cdot\vec{p}\}
\exp\{i(m\vec{\dot{\xi}}+\frac{e}{2c}\vec{B}\times\vec{\xi})
 \cdot\vec{r}\}.
\end{equation}
The real-valued function $\theta(t)$ and parameters $\vec{\xi}$,
$\vec{\dot{\xi}}$ should be chosen appropriately for Eq. (\ref{g4}) to be
fulfilled. This condition allows us to cast the unitary operator $W_t$ to
the following form:
\begin{equation}\label{g8}
W_t = \exp \{i\int\limits^t L(\tau, \vec{\xi}, \vec{\dot{\xi}}) d\tau\}
\exp\{-i\vec{\xi}\cdot\vec{p}\}
\exp\{i(m \vec{\dot{\xi}}+\frac{e}{2c}\vec{B}\times\vec{\xi})
 \cdot\vec{r}\}.
\end{equation}
It is sufficient to let parameters $\vec{\xi}=\vec{\xi}(t)$ be the solutions
of the classical equations of motion:
\begin{equation}\label{g9}
m\vec{\ddot{{\xi}}} = e\vec{E}(t)+\frac{e}{c} [\vec{\dot{\xi}}
\times\vec{B}].
\end{equation}

As for $\theta(t)$, it is sufficient to choose this parameter to be equal to
the classical action of an electron in the ac electric field when the
magnetic field is also present:
\begin{equation}\label{g10}
\theta(t) = \int\limits^t L(\tau, \vec{\xi}, \vec{\dot{\xi}})d\tau
=\int\limits^t[\frac{1}{2}m\vec{\dot{\xi}}^2+e\vec{E}(\tau)+\frac{e}{c}
\vec{A}(\vec{\xi})\cdot\vec{\dot{\xi}}]d\tau.
\end{equation}

During the further consideration of the problem it is convenient to
introduce a new set of independent variables: coordinates
of the center of the cyclotron orbit  $(X_0,Y_0)$
and the relative motion $(\zeta,\eta)$ instead of
Cartesian coordinates of electrons and the corresponding momenta:
\begin{equation}\label{g11}
\begin{array}{l@{\qquad\qquad}l}
x = X_0 + \zeta & \zeta = \frac{c}{e B}P_y, \\
y = Y_0 + \eta & \eta = -\frac{c}{e B}P_x
\end{array}
\end{equation}
that satisfy well-known commutation rules:
\begin{eqnarray}\label{g12}
[\zeta, \eta] = -i\alpha^2, \qquad [X_0, Y_0] = i\alpha^2,
\nonumber \\
{} [\zeta, X_0] = [\eta, Y_0] = [\eta, X_0] = [\zeta, Y_0] = 0,
\end{eqnarray}
where $\alpha = c\hbar/(|e|B)$ is the magnetic length.

In the new variables, the operator that defines the canonical transformation
has the following form:\cite{Torres05}
\begin{equation}\label{g14}
W_t = \exp\{i\theta_1(t)\}\exp\{-i\frac{y_0 X_0}{\alpha^2}\}
\exp\{i\frac{x_0 Y_0}{\alpha^2}\}\exp\{-i\frac{x_{rel}\eta}{\alpha^2}\}
\exp\{i\frac{y_{rel}\zeta}{\alpha^2}\}.
\end{equation}
Here
\begin{equation}\label{g15}
\theta_1(t) = \theta(t)
-\frac{m\omega_c}{2}[x_{rel}y_{rel}+x_0 y_{rel}+x_0 y_0-x_{rel}y_0],
\end{equation}
where $x_{rel}$, $y_{rel}$, $x_0$, $y_0$ are relative coordinates and
coordinates of cyclotron orbit center corresponding to the displacement
vector $\vec{\xi}$.

Using the explicit expression for the operator $W_t$ (\ref{g14}), we intend
to perform the transformation (\ref{g4}). For this, obviously, it is
sufficient to know how the following operators are transformed:
\begin{equation}\label{g16}
\begin{array}{l@{\qquad\qquad}l}
\tilde{X}_0\equiv W^\dagger_t X_0 W_t = X_0+ x_0, &
\tilde{Y}_0\equiv W^\dagger_t Y_0 W_t = Y_0+ y_0,\\
\tilde{\zeta}\equiv W^\dagger_t \zeta W_t = \zeta +x_{rel}, &
\tilde{\eta}\equiv W^\dagger_t \eta W_t = \eta + y_{rel}.
\end{array}
\end{equation}

For Eq. (\ref{g4}) to hold, it is sufficient to let $x_{rel}$, $y_{rel}$, $x_0$,
$y_0$ be the solutions of the classical equations of motion
\begin{eqnarray}\label{g18}
-m\omega_c\dot{x}_0 + eE_y = 0,\qquad
-m\omega_c\dot{y}_{rel}-m\omega_c^2 x_{rel}+e E_x = 0,\nonumber\\
m\omega_c\dot{y}_0 + e E_x = 0,\qquad
m\omega_c\dot{x}_{rel}-m\omega_c^2 y_{rel} + e E_y = 0.
\end{eqnarray}
Besides that,
\begin{equation}\label{g19}
\dot{\theta}_1 = m\omega_c \dot{y}_0 x_0
+ m\omega_c \dot{x}_{rel} y_{rel}
-\frac{m\omega_c^2}{2}
\{x_{rel}^2+y_{rel}^2\} + e E_x(x_0+x_{rel}) + e E_y(y_0+y_{rel}).
\end{equation}
Then, using (\ref{g16})--(\ref{g19}), we obtain:
\begin{equation}\label{g17}
W^\dagger_t\{i\frac{\partial}{\partial t}-\frac{m\omega_c^2}{2}
[\zeta^2+\eta^2]+e E_x(X_0+\zeta) + e E_y(Y_0+\eta)\} W_t =
\{i\frac{\partial}{\partial t}-\frac{m\omega_c^2}{2}
[\zeta^2+\eta^2]\}.
\end{equation}

As the result of the canonical transformation, the Hamiltonian of the system
under consideration can be written as follows:
\begin{eqnarray}\label{g30}
\tilde{H}(t) \equiv W_t H(t) W^\dagger_t = \tilde{H}_0 +
\tilde{H}^0_{ef}\\
\tilde{H}_0 = \tilde{H}_e + H_v + \tilde{H}_{ev}(t)
\end{eqnarray}

It is essential that the Hamiltonian of electron-impurity interaction
acquires time dependence due to the canonical transformation.

\section{Non-equilibrium statistical operator}\label{linresp}

Let's consider that, before the weak measurement field is turned on, the
system, due to absorption of the microwave radiation, was in the state
described by the distribution $\rho^0(t,0)$. If the external additional weak
measurement field is applied to the system, a new non-equilibrium state will
form in it, that should be described by an extended set of the basis
operators. The new state is defined by the non-equilibrium statistical
operator $\rho(t,0)$. We shall apply the theory of linear response of
a non-equilibrium system to a weak measurement field in order to find this
state. This theory behaves correctly in the limiting case of slightly
non-equilibrium systems \cite{Kalashnikov71}.

Let's shortly review the theory of linear response of a non-equilibrium
system to a weak measurement field. Let the non-equilibrium system with the
Hamiltonian $H$ be acted upon by an additional weak field
\begin{equation}\label{n2}
  H_1(t)=-AF(t),
\end{equation}
where $A$ is some operator, $F(t)$ is the magnitude of the external field (a
C-number). Under this perturbation, a new non-equilibrium state is formed in
the system, and it, obviously, cannot be described in terms of the original
set of basis operators. The Liouville equation that is satisfied by the new
non-equilibrium distribution $\rho(t)$, can be represented in this form:
\begin{equation}\label{n3}
\frac{\partial\rho(t,0)}{\partial t}+(i L+i L_1(t))\rho(t,0)
=-\varepsilon(\rho(t,0)-\rho^0(t,0)), \qquad
(\varepsilon\to+0),
\end{equation}
$$ i L A=\frac{1}{i\hbar}[A,H],\qquad
i L_1(t)A=\frac{1}{i\hbar}[A,H_1(t)].$$

Here $\rho^0(t)$ is the statistical operator describing the initial
non-equilibrium state of the system with the Hamiltonian $H$.

One may treat as the initial condition for $\rho(t)$ the requirement that
$\rho(t)$ coincides with the original non-equilibrium distribution
$\rho^0(t)$ at $t\to-\infty$. Limiting our consideration to the linear terms
with respect to the small correction $H_1(t)$ to the Hamiltonian, we obtain
the following expression for $\rho(t,0)$:
\begin{eqnarray}\label{n4}
\rho(t,0)=\rho^0(t,0)-\int\limits_{-\infty}^0d t_1
e^{\varepsilon t_1} i L_1 \rho^0(t+t_1,t_1),
\end{eqnarray}
where
\begin{eqnarray}\label{n5}
\rho^0(t,0) = \varepsilon\int\limits_{-\infty}^0 d t_1
e^{\varepsilon t_1} i L_1 \rho_q(t+t_1,0).
\end{eqnarray}

The linear admittance, corresponding to an arbitrary operator $B$ in the case
when the external force obeys the harmonic law with the frequency $\omega$
can be represented as
\begin{equation}\label{n6}
\chi_{BA}(t,\omega)=-\int\limits_{-\infty}^0 d t_1 e^{(\varepsilon-i\omega)t_1}
\frac{1}{i\hbar}\Sp \{ B,e^{i t_1 L}[A, \rho^0(t+t_1,0)]\}
\end{equation}
The problem of obtaining the non-equilibrium admittance can be reduced to
calculation of the transport matrix or Green's function. Using the identity
\begin{equation}\label{n7}
\varepsilon\int\limits_{-\infty}^0 d t_1 e^{\varepsilon t_1}
e^{i t_1 L} [A, \rho^0(t+t_1,0)]=
\int\limits_{-\infty}^0 d t_1 e^{\varepsilon t_1}
e^{i t_1 L} [\dot{A}, \rho^0(t+t_1,0)]-[A, \rho^0(t)],
\end{equation}
$\dot{A}=(i\hbar)^{-1} [A, H]$, and introducing the definition of
correlation functions
\begin{equation}\label{n8}
\langle B,A \rangle=-\frac{1}{i\hbar} \int\limits_{-\infty}^0 d t_1
e^{\varepsilon t_1}\Sp\{ B e^{i t_1 L} [A,\rho^0(t+t_1,0)]\},
\end{equation}
\begin{equation}\label{n9}
\langle B,A \rangle^\omega=-\frac{1}{i\hbar} \int\limits_{-\infty}^0 d t_1
e^{(\varepsilon -i\omega)t_1} \int\limits_{-\infty}^0 d t_2
e^{\varepsilon t_2}
\Sp \{ B e^{i (t_1+t_2) L} [A, \rho^0(t+t_1+t_2,0)]\},
\end{equation}
we transform the expression for the admittance. After simple calculations,
we obtain
\begin{equation}\label{n10}
\chi_{BA}(t,\omega)=\chi_{BA}(t,0) \frac{T_{BA}(t,\omega)+\varepsilon}{T_{BA}
(t,\omega)+\varepsilon-i\omega}.
\end{equation}
\begin{equation}\label{n11}
\chi_{BA}(t,0)=\langle B,A \rangle,\qquad
T_{BA}=\frac{1}{\langle B,A \rangle^{\omega}}
\langle B,\dot{A}\rangle^{\omega}.
\end{equation}
Note that the transport matrix $T_{BA}(t,\omega)$ plays in the
non-equilibrium case exactly the same role as for the response of an
equilibrium system. The transport matrix $T_{BA}(t\omega)$ and the Green's
function $G_{BA}(t,\omega)$ are bound with the following relation:
\begin{eqnarray}\label{n12}
G_{BA}(t,\omega) \{ T_{BA}(t,\omega)+\varepsilon-i\omega \}^{-1},
\nonumber\\
G_{BA}(t,\omega) = \frac{1}{\langle B A \rangle}
\langle BA \rangle^{\omega}.
\end{eqnarray}

Thus, within the approach outlined above, the problem of non-equilibrium
admittance calculation is indeed reduced to obtaining the transport matrix
or Green's function, which, in turn, requires the technique of projection
operators to be applied.

Let's apply the approach described above for calculation of static
conductivity of non-equilibrium electrons interacting with the subsystem of
impurity centers. In this case
\begin{equation}\label{n13}
B = \frac {e\tilde{\Pi}}{m},\qquad A = e\tilde{X}^\dagger,
\end{equation}
where $\tilde{\Pi}$ is the vector column built from operators-components of
the total momentum of electrons, $\tilde X$ is the vector column with the
following components:
$\tilde{X}^\alpha = \sum_j \tilde{X}_j^\alpha,$
$X^\alpha_j$ is the $\alpha$-projection of the $j$-th electron's coordinate.

Now we introduce the projection operator $\mathcal{P}$ onto the basis
operator set $\tilde{\Pi}^\dagger$
\begin{equation}\label{n14}
\mathcal{P}\tilde{\Pi}^\dagger=\tilde{\Pi}^\dagger
\frac{1}{\langle\tilde{\Pi},\tilde{\Pi}^\dagger\rangle}
\langle\tilde{\Pi},\tilde{\Pi}^\dagger\rangle,\quad
\mathcal{P} \tilde{\Pi}=\langle\tilde{\Pi},\tilde{\Pi}^\dagger\rangle
\frac{1}{\langle\tilde{\Pi},\tilde{\Pi}^\dagger\rangle}\tilde{\Pi},
\end{equation}
where
$$
  \mathcal{P} \tilde{\Pi}^\dagger = \tilde{\Pi}^\dagger,\quad
  \mathcal{Q} = (1-\mathcal{P}),\quad
  \mathcal{P}\mathcal{Q}\tilde{\Pi}^\dagger = 0.
$$
Let's consider the identity
\begin{equation}\label{n15}
  i(L-E)\tilde{\Pi}^\dagger(E) = \tilde{\Pi}^\dagger,\quad i E =
  i\omega - \varepsilon.
\end{equation}
Now we act upon both left and right hand sides of this identity with
operators $\mathcal{P}$, $\mathcal{Q}$ in turn. Taking into account that
$$
\mathcal{P}\tilde{\Pi}^\dagger(E) = \tilde{\Pi}^\dagger G(t,\omega),\qquad
\mathcal{P} (-i E+\mathcal{Q} i L)^{-1}\mathcal{Q} i L \tilde{\Pi}^\dagger(E) = 0,
$$
after some transformations we obtain the following equation for the Green's
function:
\begin{equation}\label{n16}
G(t,\omega) = [ R(t,\omega) + i\Omega(t,\omega) -i E ]^{-1},
\end{equation}
where
\begin{equation}\label{n17}
i\Omega(t,\omega) = \frac{1}{\langle\tilde{\Pi},\tilde{\Pi}^\dagger\rangle}
{\langle\tilde{\Pi},\dot{\tilde{\Pi}}^\dagger\rangle},
\end{equation}
is the frequency matrix, and
\begin{equation}\label{n18}
R(t,\omega)=\frac{1}{\langle\tilde{\Pi},\tilde{\Pi}^\dagger\rangle}
{\langle\mathcal{Q} \dot{\tilde{\Pi}},(-i E+\mathcal{Q} i L)^{-1}
\mathcal{Q}\dot{\tilde{\Pi}}^\dagger\rangle}
\end{equation}
is the memory function.
\begin{equation}\label{n19}
T(t,\omega) = R(t,\omega) + i\Omega(t,\omega).
\end{equation}

Equations (\ref{n17}), (\ref{n18}) help us calculate the frequency matrix
and the memory function in cases when the non-equilibrium state of the
system is stationary and the statistical operator $\rho^0(t)$ either doesn't
depend upon time at all or depends periodically.

Thus, within the framework described above, for the
relaxation time $\tau$ of non-equilibrium
electrons' momentum, we obtain:
\begin{equation}\label{n21}
\frac{1}{\tau} = \operatorname{Re} R(0,0) = \frac{(-1)}
{\langle \tilde{\Pi}^\alpha,
\tilde{\Pi}^{\lambda}\rangle}\operatorname{Re}
\int\limits_{-\infty}^0 d t_1 e^{\varepsilon t_1}
\int\limits_{-\infty}^0 d t_2
e^{\varepsilon t_2}
\Sp \{\mathcal{Q} \dot{\tilde{\Pi}}^\alpha
e^{i(t_1+t_2)L}
\frac{1}{i\hbar}[\mathcal{Q}\dot{\tilde{\Pi}}^\lambda,
\rho^0]\}.
\end{equation}
Here $\rho^0$ is the non-equilibrium statistical operator, and $i L$ is the
Liouville evolution operator.

The formulas above are general enough and are valid for any stationary
non-equilibrium distribution. In the further text, we assume a concrete form
for the original non-equilibrium distribution. We take that it can be
characterized by temperatures of the subsystems of the crystal: let
$\beta_e=T^{-1}_k$ be the inverse temperature of the kinetic degrees of
freedom of electrons,  $\beta = T^{-1}$ be the inverse temperature of the
lattice. The initial non-equilibrium distribution is defined by the
quasi-equilibrium distribution $\rho_q(t)$ that can be generally written as
\begin{eqnarray}\label{n22}
\rho_q(t) = e^{-S(t)},\qquad
S(t) &=& \Phi(t) + \sum_n P^\dagger_n F_n(t),\nonumber\\
\Phi(t) &=&\ln\Sp\exp\{-S(t)\}.
\end{eqnarray}
Here $S(t)$ is the entropy operator, $\Phi$ is the Massieu---Planck
functional.
$P_n$, $F_n(t)$ is the set of basis operators and conjugate functions that
describe the system under consideration. Characterizing the state of the
system with the average values of the operators $\tilde{H}_e$ and $H_v$,
we write down the entropy operator as:
\begin{equation}\label{n23}
S(t) = \Phi(t) + \beta_e \tilde{H}_0 + \beta H_v.
\end{equation}

When calculating the momentum relaxation frequency of non-equilibrium
electrons, we'll limit ourselves to Born approximation. This means that while
calculating the memory function, it is sufficient to take into account only
the terms up to the second order in the electron-impurity interaction. This
is the reason why the non-equilibrium statistical operator $\rho^0(,0)$ can
be substituted with the quasi-equilibrium distribution (\ref{n22}), with the
Hamiltonians describing the electron interaction with scatterers being
deliberately omitted. Obviously, the evolution operator $i L\to i L_0$ should
be replaced with the one for non-interacting subsystems. For the relaxation
frequency, we have:
\begin{equation}\label{n25}
\frac{1}{\tau} = \frac{i}{\hbar^3 m n}
\operatorname{Re} \int\limits_{-\infty}^0 d t_1
\int \limits_{-\infty}^0 d t_2 e^{\varepsilon(t_1+t_2)}
\Sp\{[\Pi^y,\tilde{H}_{ev}]
e^{i(t_1+t_2)L_0} [ [ \Pi^y,\tilde{H}_{ev}],\rho_q]\}.
\end{equation}

Now we expand Eq. (\ref{n25}) using the explicit expression for
renormalized electron-impurity interaction $\tilde{H}_{ev}$ from
Appendix B and the Kubo identity
\begin{equation}\label{n28}
[\dot{\Pi}^y_{ev},\rho_q] =-\int\limits_0^1 d\lambda e^{-\lambda
S_0(t)} [\dot{\Pi}^y_{ev}, S_0(t)] e^{(\lambda -1) S_0(t)}.
\end{equation}
We obtain:
\begin{multline}\label{n27}
\frac{1}{\tau} = \frac{i}{\hbar^3 m n} \operatorname{Re}
\int\limits_{-\infty}^0 d t_1 \int\limits_{-\infty}^0 d t_2
\int\limits_0^1 d\lambda e^{\varepsilon(t_1+t_2)}
\sum_{\nu'\nu}\langle\dot{\Pi}^y_{(ev),\nu'\nu}
\frac{1}{i\hbar}[\dot{\Pi}^y_{ev},S_0]_{\nu\nu'}\rangle_i\times{}\\{}\times
\langle a^\dagger_{\nu'}a_\nu
a^\dagger_\nu(t_1+t_2+i\hbar\beta\lambda) a_{\nu'}
(t_1+t_2+i\hbar\beta\lambda)\rangle.
\end{multline}

Evaluating all commutators in turn, and averaging over both the impurity
and electron subsystems, we obtain for the relaxation frequency of
non-equilibrium electrons:
\begin{equation}\label{n29}
\frac{1}{\tau} = \frac{\pi \hbar}{m n} \sum_{\nu'\nu q
 l} q_y^2 N_i |G(q)|^2 |e^{iqr}_{\nu' \nu}|^2 \Phi^2_l(\Delta)
\{f(\varepsilon_\nu)-f(\varepsilon_{\nu'})\} \frac{\partial}{\partial
\varepsilon_\nu}\delta(\varepsilon_\nu-\varepsilon_{\nu'}+l\hbar\omega).
\end{equation}
Here
$\langle a^\dagger_\nu a_\nu\rangle=
f(\varepsilon_\nu)=\{\exp[\beta_e(\varepsilon_\nu-\zeta)]\}^{-1}$ is the
electron distribution function, $\langle\rho_q\rho_{-q}\rangle_i=N_i$
is the concentration of scattering impurity centers. See the explicit form
of $\Phi_l$ in Appendix B.

Using the expression for the relaxation frequency, we write down the formula
for diagonal components of the conductivity tensor:
\begin{equation}
\sigma_{x x}= \frac{n e^2}{m} \frac{\tau^{-1}}{\omega_c^2 + \tau^{-2}}
\end{equation}
In the situation when $\omega_c \gg \tau^{-1}$, we have:
\begin{multline}\label{n30}
\sigma_{x x} =\frac{\pi \hbar e^2}{m^2 \omega_c^2}
\sum_{\nu' \nu q l}q_y^2 N_i\int d \mathcal{E}
|G(q)|^2 |e^{iqr}_{\nu' \nu}|^2 \Phi^2_l(\Delta)
\{f(\mathcal{E})-f(\mathcal{E}+l\hbar\omega)\} \times{}\\{}
\times\frac{\partial}{\partial \mathcal{E}}
\delta(\mathcal{E}-\varepsilon_{\nu'}+l\hbar\omega)
\delta(\mathcal{E}-\varepsilon_\nu).
\end{multline}

Eq. (\ref{n30}) defines the response of a non-equilibrium system of 2D
charge carriers to a weak measurement field.

We apply Eq. (\ref{n30}) for analysis of the conductivity tensor component
$\sigma_{x x}$, assuming that the main mechanism of 2D electron scattering is
due to neutral impurities with short-range (delta) potential,
$G(q)=G=const$.
Also, for simplification of numerical calculations, we consider only
one-photon processes. This approximation is valid for intermediate values of
microwave radiation power, for which $|\Delta| \ll 1$.
In this case, one can substitute $J_0 ( | \Delta | ) \approx 1$, $J_{\pm 1}
( | \Delta | ) \approx \pm | \Delta | / 2$, and neglect the contribution of
terms with $|l| > 1$.

Using the known wave functions of the electron in the magnetic field, we
obtain:
\begin{equation}\label{n31}
  | \langle n k_x |e^{i q r} | n' k_x' \rangle |^2 = \delta_{ k_x -q_x,
  k_x'} \exp ( - \frac{\alpha^2 q^2}{2} ) \left\{\begin{array}{l}
    \frac{n!}{n' !} ( \frac{\alpha^2 q^2}{2} )^{n' - n} ( L_n^{n' - n} (
    \frac{\alpha^2 q^2}{2} ) )^2,\ n' \geq n\\
    \frac{n' !}{n!} ( \frac{\alpha^2 q^2}{2} )^{n - n'} ( L_{n'}^{n - n'} (
    \frac{\alpha^2 q^2}{2} ) )^2,\  n' \leq n
  \end{array}\right.,
\end{equation}
where $L_m^{n - m} $ is the generalized Laguerre polynomial.

The further task of calculating the diagonal components of the conductivity
tensor is to convert sums to integrals in Eq. (\ref{n30}) and to calculate
those integrals in turn. In the integral on $q$, it is convenient to change
the variables from Cartesian to polar ones, and to use the explicit
expression for $\Phi_l ( | \Delta | )$. The integral on $q^2$ is then taken
using the recurrent relation and the orthogonality relation
between generalized Laguerre polynomials. In the result, we have:
\begin{multline}\label{n34}
  \int\limits_0^{\infty}
  d ( q^2 ) q^4 e^{- \frac{\alpha^2 q^2}{2}} ( \frac{\alpha^2 q^2}{2}
  )^{|n' - n|} ( L_{\min ( n, n' )}^{|n' - n|} ( \frac{\alpha^2 q^2}{2} ) )^2
  ={}\\{}=
  \frac{8}{\alpha^6}  \frac{( \max ( n, n' ) ) !}{( \min ( n, n' ) ) !} (
  n^2 + n'^2 + 3 ( n + n' ) + 4 n n' + 2 ).
\end{multline}

Removal of singularities caused by delta functions in the density
of states in Eq. (\ref{n30}) is performed, as usual, by taking
broadening of Landau levels into account. Within the
self-consistent Born approximation, the density of states of the
$n$-th Landau level has a Gaussian form:
\begin{equation}\label{n35}
D_n(\mathcal{E}) = \left(\frac{\pi}{2 \Gamma_n^2}\right)^{1/2}
\exp \{-(\mathcal{E}-\mathcal{E}_n)^2 /(2\Gamma^2_n) \},\qquad
\Gamma^2_n = \frac{2 \gamma_n \hbar^2 \omega_c}{\pi \tau},
\end{equation}
where $\tau$ is the relaxation time determined from the mobility in zero
magnetic field and $\gamma_n=1$ for short-range scattering potential.

Integration over the energy, finally, yields:
\begin{multline}\label{n36}
  \int\limits_{-\infty}^{\infty}
  d\mathcal{E}
  (f(\mathcal{E}) - f(\mathcal{E} + l \hbar \omega))
  \frac{\partial}{\partial \mathcal{E}} D_{n'}
  (\mathcal{E}+ l \hbar \omega ) D_n (\mathcal{E}) \simeq{}\\
  - \frac{\pi^{3 / 2} (n - n') \hbar \omega_c + l \hbar \omega }
  {4 \Gamma^3} (f(\mathcal{E}_n) - f(\mathcal{E}_{n'}))
  \exp(- \frac{((n - n') \hbar \omega_c + l \hbar \omega)^2}{4 \Gamma^2}),
\end{multline}

It is convenient to write the final expression for the diagonal components
of the conductivity tensor in the following form:

\begin{equation}\label{n37}
\sigma_{x x} = \sum\limits_{n n'} K_1 K_2 K_3 K_4 K_5,
\end{equation}
where
\begin{eqnarray}
K_1 &=& \frac{\hbar e^2 N_i |G|^2}{\pi m^2 \omega_c^2 \alpha^8}
    \frac{e^2 E^2}{4 m^2 \omega^2 ((\omega^2 - \omega_c^2)^2
    + (\omega \Gamma / \hbar)^2)} \nonumber\\
K_2 &=& \left( 1 + \frac{|e_x|^2 - |e_y|^2}{2}\right) \omega_c^2
    + \left( 1 - \frac{|e_x|^2 - |e_y|^2}{2}\right) \omega^2
    - 4 \omega \omega_c \operatorname{Im} (e_x^* e_y) \nonumber\\
K_3 &=& n^2 +n'^2 + 3 (n + n') + 4 n n' + 2 \nonumber\\
K_4 &=& - \frac{\pi^{3 / 2} ((n - n') \hbar \omega_c + l \hbar\omega}
    {4 \Gamma^3}
    \exp(- \frac{((n - n') \hbar \omega_c + l \hbar \omega)^2}
    {4 \Gamma^2}) \nonumber\\
K_5 &=& f(\mathcal{E}_n) - f(\mathcal{E}_{n'})
\end{eqnarray}

\section{Numerical analysis}\label{numanal}

The numerical calculations according to eq. (\ref{n37})
were done for the following parameters:
$m = 0.067 m_0$ ($m_0$ is the free electron mass),
Fermi energy is $\mathcal{E}_F = 10\ \mathrm{meV}$,
the mobility of 2D electrons is $\mu= 0.1-1.0\times 10^7\ \mathrm{cm^2/Vs}$,
electron concentration is $n=3\times10^{11}\ \mathrm{cm^{-2}}$.
The frequency of microwave radiation is  $f= 50 - 100\ \mathrm{GHz}$,
the temperature is $T = 0.5-2.5\ \mathrm{K}$.
The magnetic field is varied between $0.02-0.3\ \mathrm{T}$.

The photoconductivity dependency on the $\omega / \omega_c$ ratio for
different values of electron mobility at the 50 GHz radiation frequency is
presented on Fig. 1. It follows from the figure that, when the electron
mobility is low ($\mu = 0.1 \times 10^7\ \mathrm{cm^2/Vs}$) the harmonics of
the cyclotron resonance have a low amplitude and because of that are not
observed. With higher electron mobility, those harmonics start manifesting
themselves, and the photoconductivity dependency on the magnetic field
acquires oscillating character (as observed in experiments
\cite{Zudov03,Mani02}). The oscillation nodes are at integer and
half-integer values of $\omega / \omega_c$, maxima are at $\omega / \omega_c
= j - \delta$, minima are at $\omega / \omega_c = j + \delta$ where $j$
takes integer values, and $\delta$ depends on electron mobility. If
$\mu = 0.3 \times 10^7\ \mathrm{cm^2/Vs}$, then $\delta = 0.25$, in
agreement with \cite{Zudov03,Mani02}. When one increases electron mobility,
$\delta$ is decreased.

The dependency of the photoconductivity of the 2DEG upon the
$\omega/\omega_c$ ratio for microwave radiation frequency equal to 150 GHz
is presented on Fig. 2. It can be seen that the demanded mobility of the
carriers in the 2DEG can be lowered by raising the frequency of the
microwave radiation.

The $K_2$ factor in Eq. (\ref{n37}) depends upon the polarization of the
microwave radiation. The photoconductivity of the 2DEG is plotted vs the
$\omega/\omega_c$ ratio on Fig. 3 for left and right circular polarizations,
as well as for longitudinal (with respect to the dc measurement electric
field) and transverse linear polarizations.

\section{Conclusions}

It has been shown that the method of non-equilibrium statistical operator
together with the canonical transformation of the Hamiltonian allows one to
put forward a theory of the non-equilibrium 2DEG linear response to a weak
measurement electric field. The resulting theory describes the dependency of
the 2DEG magnetoresistance upon electron mobility, magnetic field and
microwave radiation frequency, observed in the experiments. This theory also
predicts the possibility of the oscillating photoconductivity observation in
2DEG mobility lower than the one used in experiments \cite{Zudov03, Mani02},
if one uses elevated microwave radiation frequencies and left circular
polarization.

\newpage

\section*{A. Equations of motion. Floquet states}

If the electric field is
$\vec{E}(t) = \vec{E}_0 \operatorname{Re}\{\vec{e}\ e^{i\omega t}\}$
($\vec{e}$ is a complex-valued vector that describes the polarization of the
microwave radiation, $E_0$  is the amplitude of the ac electric field), then
the solution of the classical equations of motion can be written down as
\begin{eqnarray}\label{p8}
x_{rel} = \frac{e E_0}{m\omega_c}\operatorname{Re}\left(
  \frac{e_x \omega_c +i e_y \omega}{\omega^2-\omega_c^2}
  e^{i\omega t}\right),\qquad\qquad
x_0 = \frac{e E_0}{m\omega_c\omega}\operatorname{Re}(ie_y e^{i\omega t}),
\nonumber\\
y_{rel} = \frac{e E_0}{m\omega_c}\operatorname{Re}\left(
  \frac{e_y \omega_c -i e_x \omega}{\omega^2-\omega_c^2}
  e^{i\omega t}\right),\qquad\qquad
y_0 = -\frac{e E_0}{m\omega\omega_c}\operatorname{Re}(i e_x e^{i\omega t}).
\end{eqnarray}

According to Floquet theorem, the wave function of the electron can be
represented in the form
$\Psi(t)=exp(-i E_{\mu}t) \Phi_{\mu}(t)$, where
$\Phi_\mu(t+\tau_\omega)=\Phi_\mu(t)$ is a periodic function of time, and
for the quasienergies the following expression is true:
\begin{eqnarray}\label{p9}
E_\mu & = & E_\mu^{(0)}+ E_\omega \nonumber
\\
E_\mu^{(0)} & = & \hbar\omega_c (\mu+\frac{1}{2}),\qquad\qquad
E_\omega = \frac{e^2 E^2_0 [1+2\omega_c\operatorname{Im}(e^*_x e_y)/\omega ]}
{4m (\omega^2 - \omega_c^2)},
\end{eqnarray}
where $E_\mu^{(0)}$ is the energy corresponding to a Landau level and
$E_\omega$ is the energy shift caused by microwave radiation.

\section*{B. Transformation $W(t)H_{ev}W^\dagger(t)$}

It follows from the structure of the electron-impurity interaction
Hamiltonian that its canonical transformation reduces to the transformation
of the operator exponent
$$e^{i\vec{q}\vec{r}}=e^{i\{q_x(X_0+\zeta)+q_y(Y_0+\eta)\}}.$$
Taking the explicit expression (\ref{g14}) for the canonical transformation
operator into account, we obtain:
\begin{equation}\label{p1}
W(t) e^{i\vec{q} \vec{r}} W^\dagger(t) =
e^{i\vec{q}\vec{r}} e^{-i \{q_x(x_0+x_{rel})-iq_y(y_0+y_{rel})\} }.
\end{equation}

Using expressions (\ref{p8}), we have:
\begin{equation}\label{p3}
 \exp(-i \{q_x(x_0+x_{rel})-iq_y(y_0+y_{rel})\}) =
 \exp(-i\operatorname{Re} [\Delta e^{i\omega t}]),
\end{equation}
where
\begin{eqnarray}\label{p4}
 \Delta = \frac{e E}{m \omega (\omega_c^2 - \omega^2)}
\left( \omega\left(q_x e_x + q_y e_y\right) + i \omega_c \left(q_x
e_y - q_y e_x\right) \right).
\end{eqnarray}

Using the well-known expansion of the exponent in terms of the Bessel
functions $J_l$, one can classify the processes by the number of
participating photons:
\begin{equation}\label{p6}
W(t) e^{i\vec{q} \vec{r}} W^\dagger(t) =
\sum\limits_{l=-\infty}^{\infty} e^{i\vec{q} \vec{r}}
\left(\frac{\Delta} {i|\Delta|} e^{i\omega t} \right)^l J_l(|\Delta|).
\end{equation}

Thus, as a result of the canonical transformation, the electron-impurity
interaction Hamiltonian can be written as:
\begin{equation}\label{p7}
\tilde{H}_{ev}(t) = \sum_q
G(q) \rho_{-q} e^{i\vec{q} \vec{r}}
\sum\limits_{l=-\infty}^{\infty}
\left(\frac{\Delta} {i|\Delta|}
e^{i\omega t}\right)^l J_l(|\Delta|)
 \equiv\sum_{ql} G(q) \rho_{-q} e^{i\vec{q} \vec{r}}\Phi_l,
\end{equation}
where $G(q)$ is the Fourier transform of the potential of the
electron-impurity interaction.

\newpage

\begin{figure}
\center\includegraphics{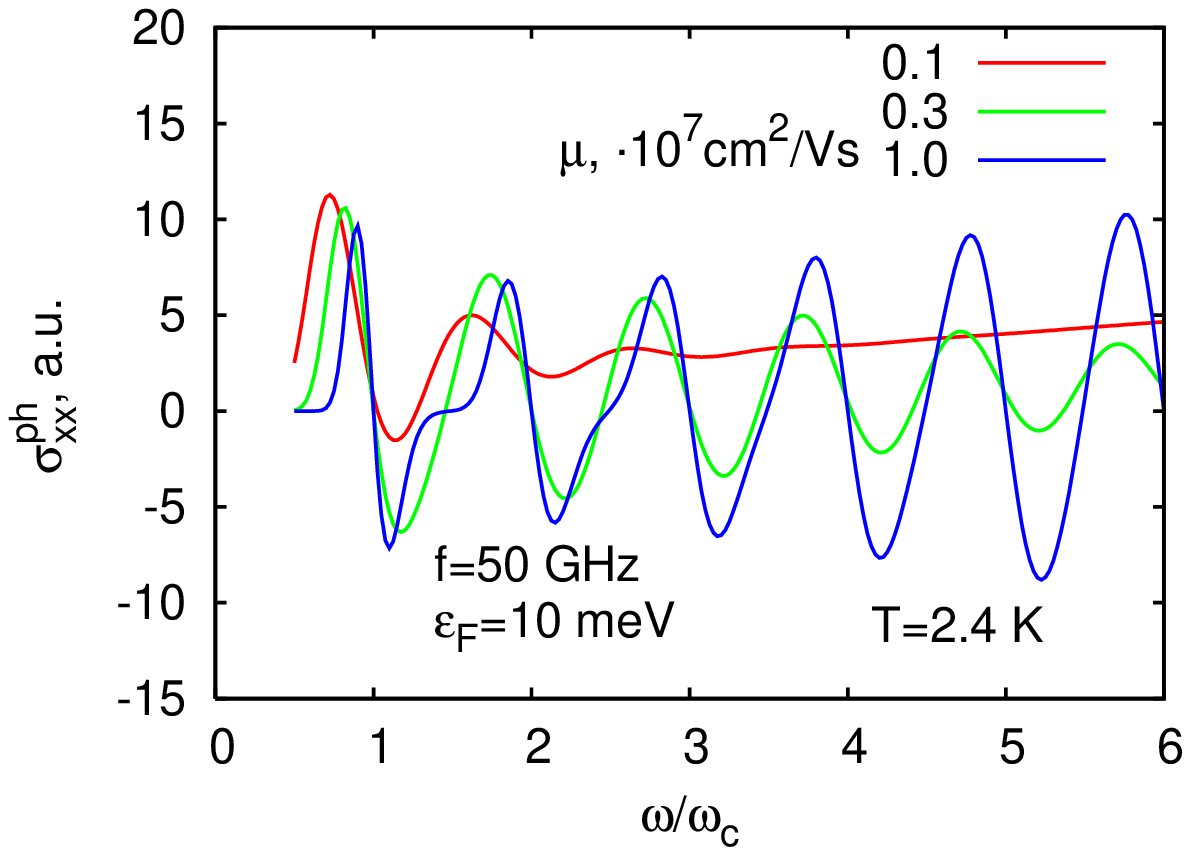} \caption{Dependency of the 2DEG
photoconductivity upon the $\omega/\omega_c$ ratio. The microwave
radiation frequency is 50 GHz.}
\end{figure}

\begin{figure}
\center\includegraphics{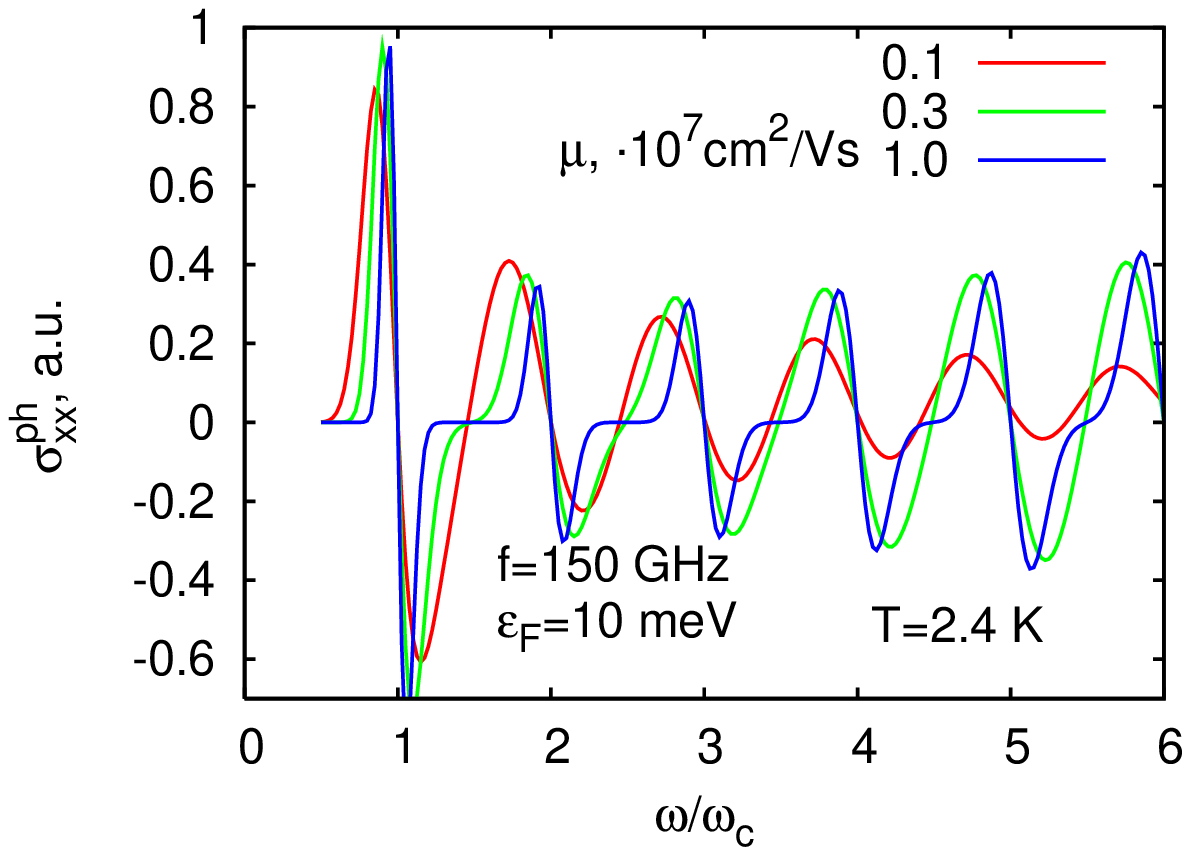} \caption{Dependency of the
2DEG photoconductivity upon the $\omega/\omega_c$ ratio. The
microwave radiation frequency is 150 GHz.}
\end{figure}

\begin{figure}
\center\includegraphics{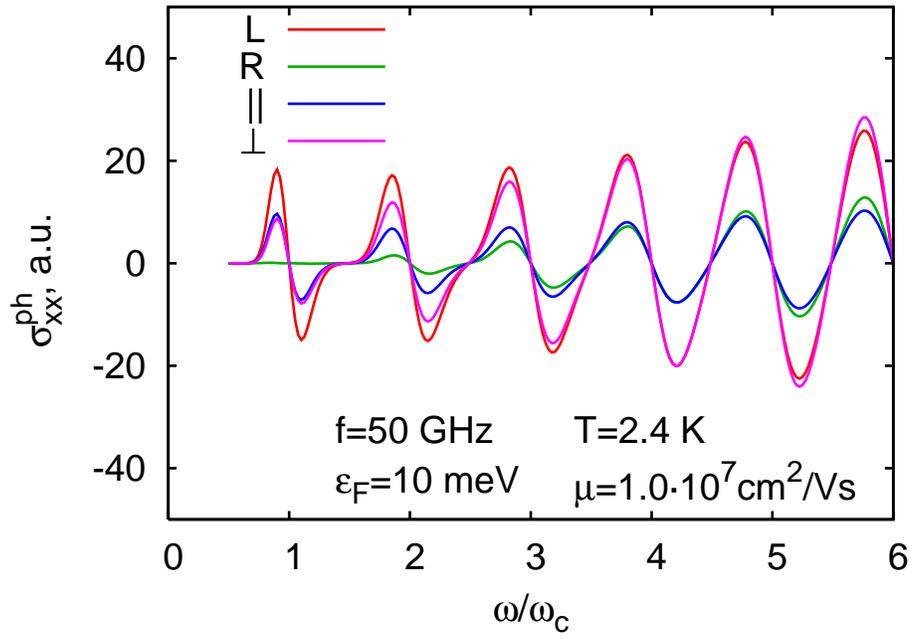} \caption{Dependency of the 2DEG
photoconductivity upon the $\omega/\omega_c$ ratio for various
polarizations of the microwave radiation: left circular (L), right
circular (R), linear longitudinal with respect to the dc field
($||$), linear transverse with respect to the dc field ($\perp$).}
\end{figure}
\end{document}